\documentclass[12pt,onecolumn]{article}
\usepackage{amssymb,amsfonts,amsmath}
\usepackage{algorithmic}
\usepackage{array}
\usepackage{mdwmath}
\usepackage{mdwtab}
\usepackage{fixltx2e}
\usepackage{amsthm}
\usepackage{setspace}
\usepackage{latexsym}
\usepackage{mathtools}
\usepackage{graphics}
\usepackage{graphicx}
\usepackage{epsfig}
\usepackage{color}
\usepackage{authblk}
\usepackage{lineno}
\usepackage{booktabs}
\usepackage{hyperref}
\usepackage{natbib}

\usepackage[margin=20mm]{geometry}

\begin{document}

\title{Macroscopic cortical dynamics: Spatially uncorrelated but temporally coherent rich-club organisations in source-space resting-state EEG}
\author[a,b,c]{\small{Steve Mehrkanoon}\thanks{Corresponding author, Email: steve.mehrkanoon@unimelb.edu.au}
}
\affil[a]{Department of Electrical and Electronic Engineering, The University of Melbourne, Melbourne, Australia}
\affil[b]{The Florey Institute of Neuroscience and Mental Health, Melbourne, Australia}
\affil[c]{Department of Neuroscience, Monash University, Melbourne, Australia}

\date{}
\maketitle
\hspace{3cm} Running title: \textbf{Macroscopic neuronal population dynamics}
\doublespace
\newpage
\begin{abstract}
Synchronous oscillations of neuronal populations support resting-state cortical activity. Recent studies indicate that resting-state functional connectivity is not static, but exhibits complex dynamics. The mechanisms underlying the complex dynamics of cortical activity have not been well characterised. Here, we directly apply singular value decomposition (SVD) in source-reconstructed electroencephalography (EEG) in order to characterise the dynamics of spatiotemporal patterns of resting-state functional connectivity. We found that changes in resting-state functional connectivity were associated with distinct complex topological features, ``Rich-Club organisation'', of the default mode network, salience network, and motor network. Rich-club topology of the salience network revealed greater functional connectivity between ventrolateral prefrontal cortex and anterior insula, whereas Rich-club topologies of the default mode networks revealed bilateral functional connectivity between fronto-parietal and posterior cortices. Spectral analysis of the dynamics underlying Rich-club organisations of these source-space network patterns revealed that resting-state cortical activity exhibit distinct dynamical regimes whose intrinsic expressions contain fast oscillations in the alpha-beta band and with the envelope-signal in the timescale of $<0.1$ Hz. Our findings thus demonstrated that multivariate eigen-decomposition of source-reconstructed EEG is a reliable computational technique to explore how dynamics of spatiotemporal features of the resting-state cortical activity occur that oscillate at distinct frequencies.\\
\\
\textbf{keywords}$-$Dynamic connectivity, Resting-state EEG, source-space analysis, Graph theoretical metrics
\end{abstract}
\newpage
\section{Introduction}
The human brain is a hierarchical complex network composed of many interacting parts (network nodes) always active and evolve dynamically $-$ the human connectome \citep{Bullmore2009186, Fornito2013426}. Brain networks are thought to provide the anatomical basis for information processing and cognitive representation \citep{Honey2007fc, Sporns2004418, Bullmore2009186, Fornito2013426}. Network theory has been used to investigate the structural connectivity of the brain (wiring diagram) and its functional connectivity (FC) $-$ the temporal correlation structure of spatially distributed cortical activities \citep{Tononi2003, Horwitz2003466}. Functional connectivity analysis of resting-state functional magnetic resonance imaging (fMRI) data has become a widely used means of investigating the organisation of intrinsic brain activity \citep{Beckmann20051001}. Multivariate analyses of these data have revealed coherent spontaneous fluctuations across widely distributed brain areas that define specific resting-state networks with high consistency across subjects, which mirror the organisation of cognitively salience networks (SN) \citep{Biswal1995537, Smith200913040, Damoiseaux200613848}. A constellation of these brain regions show higher metabolic activity at rest than during cognitive tasks and have been labeled the Default Mode Network (DMN) \citep{Raichle2001676, Greicius2003253}. The DMN is anti-correlated with task-positive networks that are activated during specific cognitive tasks including visual, sensorimotor, attentional and executive systems \citep{Fox20059673}. It has been shown that the functional organisation of the brain is related to the widely studied resting-state networks such as the DMN \citep{Raichle2001676, Greicius2003253} and SN \citep{Biswal1995537, Smith200913040}, however, there is not such gold standard to map a complete relationship between these widely distributed resting-state networks. The DMN has been shown to exhibit higher large-scale neural activity during resting-state in order to receive neural information from widely distributed brain regions \citep{Liao201057}, whereas SN is thought to be central during switching (i.e. activating and deactivating) distinct cortical networks \citep{Sridharan200812569}. Interestingly, patterns of resting-state FC have been shown to resemble those identified by task-based data \citep{Fox200610046, Vincent200783, Smith200913040}.\\
\\
Recent neuroimaging studies have reported that FC is time-varying. Dynamic changes in FC have been observed in magnetoencephalography (MEG)\citep{Gross2001694} and recent fMRI studies. For example, coherent blood oxygenate level dependent (BOLD) activity is modulated by learning \citep{Bassett200619518}, cognitive and affective states \citep{Ekman201216714, Allen2012603} and also spontaneously \citep{Kitzbichler2009, Britz20101162, Chang201081}. More recently, fMRI studies have confirmed dynamic changes in cortical networks \citep{Chang201081, Kang20111222, Leonardi2013937, Hutchison2013360, Allen2014663, Di201337}. Emerging evidence from fMRI studies thus suggests that indices sensitive to dynamic FC metrics may capture changes in macroscopic neural activity patterns underlying critical aspects of brain functions, although limitations with regard to analysis and interpretation remain \citep{Hutchison2013360}. In contrast to fMRI time-series, electrophysiological data capture neuronal activity at high temporal resolution ratio and are thus ideally suited to capture rapid changes in cortical networks. Detailed analysis of dynamic changes in cortical networks thus has the potential to reveal novel insights into the mechanisms that organise macroscopic neural activity and its functional roles as the brain adapts to a changing environment. The main drawback of the estimation of FC directly from the scalp electroencephalography (EEG) is the dependence of connectivity patterns to the changes of the position of a common reference-electrode. That is, changing the position of a common reference-electrode leads to changes in the patterns of FC derived from EEG signals \citep{Brunet2011}. However statistically significant connectivity patterns derived form multivariate decomposition of the joint time-frequency interdependence between EEG signals have shown fairly robust networks \citep{Mehrkanoon2014338}. Thus, to overcome the problem of volume conduction and the effect of the reference electrode, source reconstruction technique has been previously proposed \citep{PascaulM1993532}. Source reconstruction involves inverting forward models that map source electrical fields into the EEG lead field \citep{Brunet2011, Mehrkanoon2014BC}.\\
\\
Given the emerging evidence of dynamic FC at rest and its crucial role for characterising the functional organisation of the brain, the aim of our study was to examine whether changes in resting-state network topographies are associated with variations in the connectivity configuration of the DMN, SN, and motor network. We developed a source-space multivariate resting-state EEG data-driven approach in order to characterise spatiotemporal coherent fluctuations of the whole-brain FC. Prior works in this very active field have generally focussed linear correlations between band-limited power fluctuations, i.e. amplitude synchronisation \citep{Lewis200917558, Hipp2012, Jin2013}. We extend beyond bivariate analyses to present a complete multivariate decomposition. In this way, we derive the principal resting-state network dynamics from human source-reconstructed EEG data, that is the low-dimensional orthogonal subspace of resting-state cortical activity. Applying this technique to human source-space EEG, we identify four robust resting-state network patterns that are consistently expressed across human subjects. Our approach exploited the dynamics underlying these networks. These networks showed a variety of distinct spatial topological features whose intrinsic expressions contain intrinsic phase and amplitude modes that fluctuate on fast and slow timescales, effectively spanning the space of resting-state cortical activity. 

\section{Methods and Materials}
\subsection{Data acquisition}
Ten healthy subjects (mean age 27.4 years; range 20$-$34 years; 4 females) participated after giving informed consent. The Subjects were instructed to relax with eyes closed and refrain from falling asleep. Surface EEG was acquired during a single 10-minute session using BrainAmp MR Plus amplifiers (Brain Products, Munich, Germany) and custom electrode caps (Easy Cap, Falk Minnow Services, Herrsching-Breitbrunn, Germany) arranged according to the international 10$-$20 system. Details of the processes of EEG acquisition, ICA-based artefact removal and band-pass filtering can be found in our previous works \citep{Mehrkanoon2014338, Mehrkanoon545616}. 
\subsection{Source reconstruction}
Analyses of the acquired EEG analysis were undertaken using a combination of publically available and in-house programs written in MATLAB (The Mathworks, Natick, MA). We used independent component decomposition (InfoMax ICA) to remove electro-occulograph (EOG) and electromyograph (EMG) artefacts from band-pass filtered EEG. For each participant, artefact-free EEG was down-sampled at 512 Hz before source reconstruction. A sample rate of 512 Hz was deemed sufficient to our objectives, as the spectral analyses were confined to frequencies below 80 Hz. We then employed the standard Low Resolution Electrical Tomography (sLORETA) algorithm to estimate source signals \citep{Pascual-Marqui20025} computed by a Locally Spherical Model with Anatomical Constraints (LSMAC) (\url{brainmapping.unige.ch/cartool},  \citep{Brunet2011}) and the standard MNI152 brain template for all subjects. The LSMAC model does not require the estimation of a best-fitting sphere, but instead uses the realistic head shape and local estimates for the thickness of scalp, skull and brain underneath each local electrode. A total of 5400 solution points (henceforth referred to as voxels) were regularly distributed within the gray matter of the cortex and cerebellum. The forward model was solved by an analytical solution using a 3-layer conductor model. The source-space was subdivided into 90 anatomically defined regions of interest (ROIs) (cerebelum was excluded from the source-reconstruction) according to the macroscopic anatomical parcellation of the MNI template using the Automated Anatomic Labelling (AAL) map \citep{Mazziotta20011293}. On average, each ROI consisted of about 50 voxels. From these voxels we then determined one voxel per region that has the minimum average Euclidean distance to the rest of voxels in that brain region, the so-called centroid voxel. We then applied principal component analysis (PCA) to each of the centroid voxels separately, i.e., PCA to the ``X, Y, and Z" components of a voxel's time-series, to extract the orientation of the source-dipole dynamics within each ROI. The first PCA component (i.e., the first PCA projection) estimated from a single centroid voxel time-series was then used for further analysis. This is similar to the approach described by \citep{Hipp2012, Mehrkanoon2014BC}. 
\subsection{Dynamic connectivity analysis}
We then estimated dynamic FC by computing the pairwise Pearson cross-correlation between all 90 brain regions using a sliding-window technique  \citep{Chang201081}, yielding a 90$\times$90 correlation matrix for each sliding-window data: The sliding-window cross-correlation between the time series $\{x_t\}_{t=1}^{N}$ and $\{y_t\}_{t=1}^{N}$ was given by
\begin{equation} 
r_{xy}(t,\tau)=\text{cross-corr}\Big([x(t+\tau,t+\Delta t)], [y(t,t+\Delta t)]\Big) \ , 
\end{equation} 
where $\tau$ and $\Delta t$ denote the time-lag index and window length respectively. Although we used a constant sliding-time window to estimate cross-correlation between the voxels time-series, recent fMRI studies have used different window lengths at the order of 10 s to investigate the reliability and variability of the BOLD-driven FC measures \citep{Allen2014663, Bassett20117641, Chang201081, Handwerker20121712, Hutchison2013360, Leonardi2013937}, we used a constant sliding-time window length (i.e. $\Delta t$) of 4 s to segment source-voxels time-series. Functional connectivity was then defined as the maximum cross-correlation within a time-lag of $\tau=\pm$25 ms. We also excluded zero time-lag correlation to eliminate potentially spurious correlations caused by the problem volume conduction \citep{Nolte20042292}.  Cross-correlation measures were then Fisher transformed (i.e. $z=\text{atanh}(r)$) to make them approximately normally distributed. For each subject, we constructed the whole-brain dynamic FC matrix $\mathbf{R}_{\text{s}}$ whose column entries were the vectorised time-lag correlation matrices, resulting in a $\frac{N^2-N}{2}\times T_\text{s}$, where $N$ and $T_\text{s}$ are the numbers of the voxels time-series and the sliding windows respectively. To vectorise a connectivity matrix $R_{(N\times N)}$, we kept the upper triangular part of the matrix. By concatenating subjects' dynamic FC matrices obtained from the 10 subjects, we constructed a multivariate dynamic FC matrix, resulting in a $\frac{N^2-N}{2}\times [T_\text{s} \times 10]$ matrix given by $\mathbf{X}=[\mathbf{R}_1(1) \hdots \mathbf{R}_1(T_\text{s}),\hdots,\mathbf{R}_{10}(1) \hdots \mathbf{R}_{10}(T_\text{s})]$. We first mean centred the multivariate dynamic FC matrix $\mathbf{X}$ to minimise the effects of inter-subject variability on the group-level analysis.  For each subject, the mean of each row was subtracted from the FC matrix $\mathbf{R}_i$, where $i=1,\hdots, 10$ (i.e. $\mathbf{R'}_i=\mathbf{R}_i-\bar{\mathbf{R}}_i$). Hereafter we use $\mathbf{X'}$ as the mean-centred group-level FC matrix instead of the notation $\mathbf{X}$ expressed in the above.\\
\\
To identify patterns of large-scale network connectivity across time and subjects, we decomposed the dynamic connectivity matrix $\mathbf{X'}$ using the singular value decomposition (SVD). The SVD is a multivariate statistical technique to detect principled and structured patterns that explain the differences in the collection of vectorised cross-correlation matrices. A set of patterns in an orthogonal subspace typically explain most of the dynamics present in all of the cross-correlation matrices across subjects. We hence applied SVD, which identified singular vectors and singular values of the group-level connectivity matrix $\mathbf{X'}$:
\begin{equation}
\mathbf{X'}=\mathbf{U\Sigma V^T} \ , 
\end{equation}
where the $\mathbf{U}\in\mathbb{R}^{(\frac{N^2-N}{2}\times \frac{N^2-N}{2})}$ reflect all possible pairwise eigenmodes of FC for a given $N=90$ voxels time-series, and $\mathbf{V}\in\mathbb{R}^{\left((\frac{N^2-N}{2}\times T_\text{s}) \times (\frac{N^2-N}{2}\times T_\text{s})\right)}$ renders the temporal evolutions (i.e., dynamcis) of the FC measures across time and subjects given by the columns in $\mathbf{U}$. In fact, the temporal evolutions of the FC measures reveal the time-course of FC oscillations that is associated with the most explained variance eigenconnectivity mode. The columns in each of the matrices $\mathbf{U}$ and $\mathbf{V}$ are linearly independent $-$ an eigenspace. The $\mathbf{\Sigma}$ is a pseudo-diagonal matrix (i.e. singular values as the explained variance of each eigenconnectivity pattern) whose top $P$ rows contain diag\{$\mathbf{\Sigma\}}=\{\sigma_1,\sigma_2,\hdots,\sigma_P\}$ (with ordered diagonal entries $\sigma_1\geq\sigma_2\geq\hdots\geq\sigma_P$) and whose bottom ($T_{\text{s}}-P$) rows are all zeros. We selected the first $P$ columns of the  matrices $U$ and $V$ by computing a threshold using the singular values of the connectivity matrix $\mathbf{X}'$ as $P=\frac{\sigma^2_1}{\sum_i\sigma^2_i} \ , i=2,\hdots , \frac{N^2-N}{2}$. 
\subsection{Statistical analysis}
To determine whether the network connectivity patterns are statistically significant, we compared them to surrogate-driven connectivity modes. For each subject, the surrogate data was constructed using the Fourier phase randomisation of the segmented voxels' signals \citep{Theiler199277}. Similar to the original analysis, the across-subjects surrogate data was decomposed by the SVD and the singular values were obtained. This analysis was performed for 1000 realisations to estimate the distribution of surrogate components. Network connectivity measures and their associated temporal evolutions were considered statistically significant if their singular values exceeded 99\% of the corresponding singular values of the surrogate distribution (p$<$0.05). The significance of the network edges was examined using bootstrapping approach \citep{McIntosh2004}. To this end, FC matrices of a random set of subjects were selected and decomposed identical to the original data. By randomly selecting subjects with replacements, the between-subject standard deviation was estimated from 1000 realisations. That is, whilst the mean across the 1000 surrogates rendered the original singular vectors (network connectivity) and temporal evolutions (network dynamics), the standard deviation across the 1000 surrogates was used to estimate the standard deviation across subjects \citep{efron1986bootstrap}. The singular vectors and values were converted to z-scores and the edges (voxel combinations) were considered significant if z-score $>$ 1.96 (p$<$0.05). 
\subsection{Network analysis}
Network theory provides a mathematical account to assess brain connectivity. In order to quantitatively evaluate the network structure of the significant resting-state connectivity patterns, graph metrics including the node degree, node-betweenness centrality, and small-worldness were derived \citep{Bullmore2009186, Rubinov20101059}. The degree expresses the number of weighted connections between a node and the remaining nodes of the network under analysis. The node-betweenness centrality quantifies how many of the shortest paths between all other node pairs in the network pass through any particular node \citep{Rubinov20101059}. The small-worldness quantifies brain's ability in functional segregation (information processing) and integration (combining the information processed) \citep{Rubinov20101059}. We also sought to characterise complex topological feature of resting-state networks using ``Rich-Club'' organisation, which implies that highly connected brain regions $-$ hubs of the large-scale brain network $-$ are more densely interconnected with each other than expected by chance \citep{Colizza2006110, Schroeter20155459}. The following steps were considered to compute the rich-club coefficients of the significant networks. For a given network connectivity (or graph), all nodes that showed a number of connections with degree $\leq k$ were removed from the network. For the remanning nodes and edges, the rich-club coefficient $\phi(k)$ was computed as the ratio of connections present between the remaining nodes and the total number of possible connections that would be present when the network would be fully connected. The rich-club coefficient $\phi(k)$ is given by \citep{Colizza2006110}: $\phi(k)=\frac{2E_{>k}}{N_{>k}(N_{>k}-1)}$.  
\subsection{Temporal contents of the networks} 
In addition to the network analysis, we also sought to characterise the temporal information of the networks. This was achieved by measuring the distance between the probability density functions (pdfs) of the networks' temporal evolutions using a kernel density estimation approach. We also sought to characterise the oscillatory activity underlying each significant resting-state network by estimating the spectral contents of the temporal evolutions associated with these networks.
\subsubsection{Diversity and closeness between networks}
First, we used a kernel based approach to estimate the pdf of the temporal evolution of each significant resting-state network. Kernel is a nonparametric unbiased and consistent estimator for the pdf of a given data $\mathbf{x}$, the so-called Parzen window \citep{EParzen1965}. The pdf of the $i$th network's temporal evolution $V_i$ (i.e. the $i$th right-singular vector of $\mathbf{X'}$) was estimated by the Gaussian kernel given by
\begin{equation}
\label{eq8}
\hat{f_i}_{_\alpha}=\hat{f}_\alpha(\sqrt{\sigma_i}V_i)=\frac{1}{T_\text{s}\sqrt{2\pi}\alpha}\displaystyle\sum_{t=1}^{T\text{s}}\text{exp}\Big(-\frac{(V_i-V_i(t))^{2}}{2\alpha^{2}}\Big),   
\end{equation} 
where $\hat{f_i}_{_\alpha}$ denotes the estimated pdf based on the kernel bandwidth $\alpha$. Optimum $\alpha$ can be determined by the Sliverman's rule \citep{Silverman1986}. We then used the Jensen-Shannon distance \citep{Grosse2002041905/1} between the pdfs of the networks' temporal evolutions to quantify diversity and closeness of the dynamics underlying each significant resting-state network. The total number of possible Jensen-Shannon distance measures between $M$ number of pdfs is given by $\frac{M\times(M-1)}{2}$, which are the entries to a $M\times M$ matrix. Average linkage clustering was used to reveal network hierarchy.
\subsubsection{Spectral contents of the networks}
Because cortical dynamics can be the result of interplay between slow and fast oscillations within the distributed brain networks \citep{Bullmore2009186}, we sought to characterise spectral contents of the temporal evolutions associated with the significant resting-state networks. To this end, we first decomposed the temporal evolutions of the networks to a set of time-series with distinct oscillatory activity, the so-called intrinsic mode function (IMF), using the empirical mode decomposition (EMD) approach \citep{Huang1998903}. We then estimated the time-frequency spectra of the IMFs using a complex-Morlet wavelet time-frequency approach \citep{Daubechies1988605, MehrkanoonEURASIP2013}. Power spectra of the IMFs were estimated to quantify detailed spectral contents underlying the significant resting-state networks. In addition to the spectral contents of the dynamics underlying resting-state networks, we quantified modulations of the network dynamics to reveal how network oscillations vary over time. To this end, we computed the Hilbert amplitude (i.e. the envelope signals) of the IMFs. In fact, power spectra and the Hilbert envelopes of the temporal evolutions of the networks provide detailed spectral information of network oscillations over multi timescales in resting-state cortical activity. 
\section{Results}
We applied a multivariate decomposition technique to the resting-state source-space FC measures estimated by a sliding-window approach. Out of 4005 networks (i.e. $\frac{N\times(N-1)}{2}$, where $N=90$ voxels time-series), four networks were associated with significant dynamic FC at the threshold of $p<0.05$ after statistical correction for the number of networks and edges, including the SN and DMNs.
\subsection{Spatial patterns of cortical activity}
\begin{figure}[htbp]
\centering{\includegraphics[height=19cm,width=15cm]{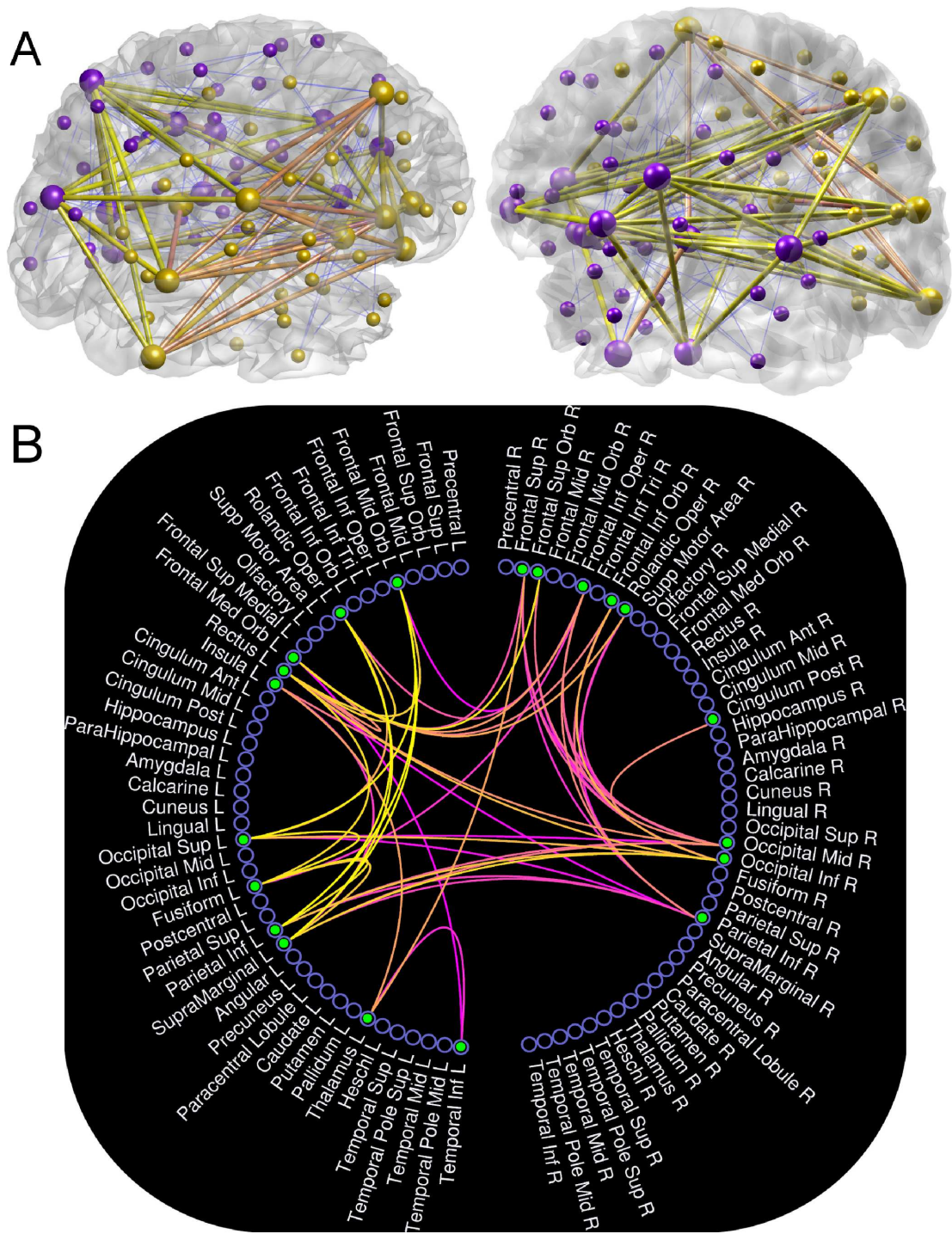}}
\caption{\textbf{Spatial topography of the rich-club organisation of the resting-state network connectivity 1.} A) Rich-club topological features of the left and right hemispheres: Bilateral FC between homologous frontal gyri, occipital and parietal lobules. The blue thin edges represent feeders (or feed edges), and B) circulogram connectome representing rich-node and rich-edges in network 1. Yellow and purple edges are the most strongest and weakest significant edges in the network. The edges depicted in yellow and purple colours are matched among the panels A and B.}
\label{fig1}
\end{figure}

\begin{figure}[htbp]
\centering{\includegraphics[height=19cm,width=15cm]{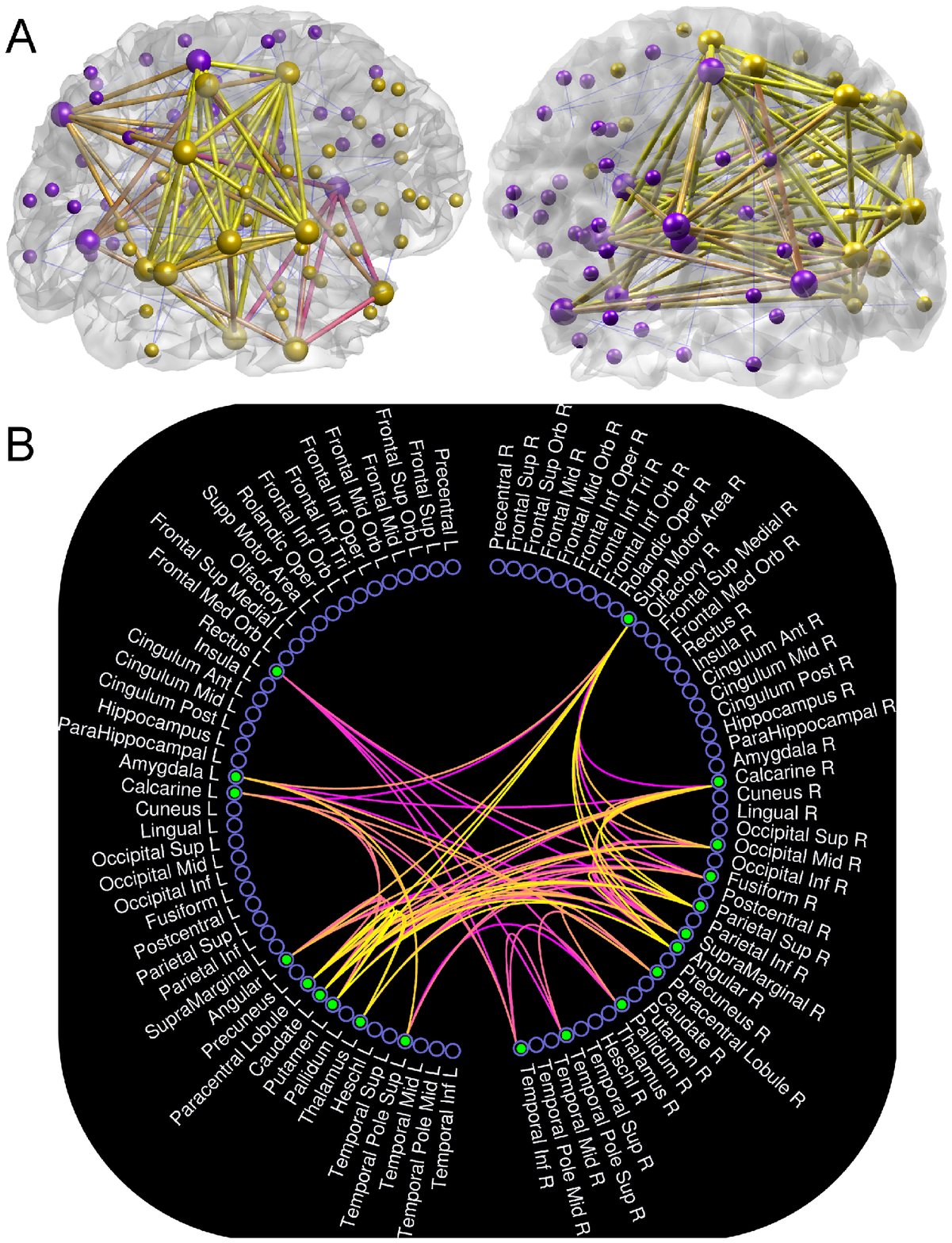}}
\caption{\textbf{Spatial topography of the rich-club organisation of the resting-state network connectivity 2.} A) Rich-club topological features of the left and right hemispheres: Functional connectivity between left ganglia gyri, right parietal areas, right supplementary motor area, and left and right temporal gyri. The blue thin edges represent feeders (or feed edges), and B) circulogram connectome representing rich-node and rich-edges in network 2. Yellow and purple edges are the most strongest and weakest significant edges in the network. The edges depicted in yellow and purple colours are matched among the panels A and B.}
\label{fig2}
\end{figure}

We thus captured spatiotemporal patterns of endogenous cortical activity, which are the four significant (p$<$0.05) resting-resting networks with distinct spatial topographies across 10 subjects (Figures \ref{fig1}$-$\ref{fig4}). These four networks mirror structure of a communication between cortical regions with variety of symmetric, regional or distributed topographies. Bilateral FC changes related to the SN were mainly located between homologous fronto-parietal regions spanned on the left and right prefrontal and parietal gyri and left insula, and the regions of the DMN, which were mainly located in homologous bilateral occipital FC and left temporal gyri (network 1). We found 5 rich nodes in network 1: Left insula, left occipital lobe, left prefrontal and parietal gyri, right frontal superior and medial orb, respectively (Figure \ref{fig1}, panels A,B). The purple and yellow nodes represent the Euclidean centers of each specific brain region: Large nodes and their connections (or edges) exhibit the rich-nodes (i.e. the green nodes in panel B) and rich-edges (panel B) of the network 1 (Figure \ref{fig1}A) respectively, forming a resting-state rich-club organisation. The edges shown in the yellow and purple colours exhibit the strongest and weakest rich-edges of the network respectively (Figure \ref{fig1}B).
\\
\\
Network 2 represents a spatially symmetric topology largely spanning on parietal and motor network regions consisting of temporal lobes, ganglia gyri (i.e. left caudate and putamen) and right supplementary gyri (Figure \ref{fig2}A). The rich-club organisation of network 2 was mainly located in the left basal-ganglia, right supplementary motor area, right parietal-posterior regions (Figure \ref{fig2}B), and the regions of the DMN which were mainly located in the right inferior temporal and occipital lobules (Figure \ref{fig2}B). Networks 3 and 4 show two spatially symmetric large-scale networks connecting anterior to posterior areas with the regions of the DMN and motor network (Figures \ref{fig3} and \ref{fig4}). They present inter- and intra-hemispheric connectivity patterns through symmetrical cliques over fronto-temporal regions, occipital lobe, insula and motor network (Figures \ref{fig3} and \ref{fig4}). Rich-club nodes and edges in the networks 3 and 4 consisted of the regions of the DMN which were mainly located in the fronto-temporal gyri and homologous bilateral occipital lobes (Figures \ref{fig3} and \ref{fig4}, panel B), and the regions of the motor network which were mainly located in the right supplementary motor area and the left putamen gyri (Figures \ref{fig3} and \ref{fig4}, panel B). Specifically, network 3 exhibited the main region of SN (i.e. left insula) that was connected to  the left prefrontal lobule (Figure \ref{fig3}B), whereas in network 4, the right supplementary motor area and the left  basal-ganglia nodes were the most strongly connected regions (Figure \ref{fig4}B). These four networks show the dynamics of the spatial patterns of large-scale cortical activity whose rich-nodes and edges can appear repeatedly across significant networks. Five specific nodes in these four resting-state networks were supplementary motor area, ganglia gyri, occipital area, prefrontal gyri and insula. This indicates that specific network connectivity configuration amongst specific nodes (or brain regions) can be responsible for shaping the brain's intrinsic dynamic functional organisation in resting-state. In the following sections, we further investigate the characteristics of the dynamics underlying these four resting-state FC patterns following graph theoretical analysis of these networks.
\begin{figure}[htbp]
\caption{\textbf{Spatial topography of the rich-club organisation of the resting-state network connectivity 3.} A) Rich-club topological features of the left and right hemispheres: Functional connectivity amongst the fronto-temporal regions and homologous occipital lobes, and right supplementary motor area and right frontal and temporal gyri. The blue thin edges represent feeders (or feed edges), and B) circulogram connectome representing rich-node and rich-edges in network 3. Yellow and purple edges are the most strongest and weakest significant edges in the network. The edges depicted in yellow and purple colours are matched among the panels A and B.}
\label{fig3}
\end{figure}
\begin{figure}[htbp]
\caption{\textbf{Spatial topography of the rich-club organisation of the resting-state network connectivity 4.} A) Rich-club topological features of the left and right hemispheres: Functional connectivity in the motor system (supp-motor area and ganglia gyri), and bilateral occipital lobes and fronto-temporal regions. The blue thin edges represent feeders (or feed edges), and B) circulogram connectome representing rich-node and rich-edges in network 4. Yellow and purple edges are the most strongest and weakest significant edges in the network. The edges depicted in yellow and purple colours are matched among the panels A and B.}
\label{fig4}
\end{figure}
\\
\noindent We measured the rich-club coefficient $\phi(k)$ of each network. To this end, we defined rich-nodes as the nodes with degree-level $k>$10, and then selected all the edges within the top (by degree-level) nodes. The estimated rich-club coefficients for all significant networks were closed to 1: 0.75 (network 1), 0.79 (network 2), 0.77 (network 3), and 0.75 (network 4), indicative of complex topological features of all 4 resting-state networks. The graph metrics node-betweenness centrality (i.e. hubs of the networks) and small-worldness (the index of a capacity for functional segregation and integration in the brain) revealed that these four networks are highly structured (Figure \ref{fig5}A-B). The hubs in networks 1 and 2 were the homologous hippocampus regions and left temporal pole gyri with the average node-degree $k=$25 and the betweenness centrality measure $B=$100 (panel A). In contrast, node-betweenness centrality measures increased to the average value of $B=150$ with lower node-degree measures of $k=10$ in the networks 3 and 4 (panel A). This change indicates the presence of different network structures in small-world network topology (panel B). These four networks together show small-world topology (i.e. $S^{w}\gg 1$), indicative of a potential capacity for parallel information processing (functional segregation) and integrating the information processed (functional integration) (panel B).\\
\\
Although different spatial topographies show distinct wiring patterns (i.e. eigenconnectivity patterns) of the resting-state cortical networks in an orthogonal subspace, investigation of the temporal evolutions associated with these FC patterns characterises the dynamics underlying resting-state cortical networks. We thus estimated the pdfs of the temporal evolutions associated with each of the significant networks by the Kernel density approach (Figure \ref{fig6}, panel A). Specifically, to test whether these pdfs were drawn from a normal distribution function such as Gaussian distribution, we measured the fourth order sample Kurtosis statistics of these pdfs. Note that a random variable with Normal distribution has a kurtosis of zero. Positive kurtosis indicates a ``peaked'' distribution and negative kurtosis indicates a ``flat'' distribution. The pdfs of the networks' temporal evolutions were drawn from a sub-Gaussian and super-Gaussian distributions with negative and positive  kurtosis values respectively: $\text{kurt}_1=-4.35$, $\text{kurt}_2=-3.72$, $\text{kurt}_3=1.84$, and $\text{kurt}_4=1.26$. This implies the existence of two distinct temporal evolutions with `flatness' and `peakedness' features of the distributions respectively (panel A). We also measured the distance between the pdfs of the networks' oscillations by the Jensen-Shannon divergence approach in order to classify the relationships between FC patterns. Networks 1 and 2 were classified into a group with the distance of $d_{\text{JS}}=20$, whereas networks 3 and 4 were classified into another group with the distance of $d_{\text{JS}}=8$ (Figure \ref{fig6}, panel B).
\begin{figure}[htbp]
\centering{\includegraphics[height=8cm,width=10cm]{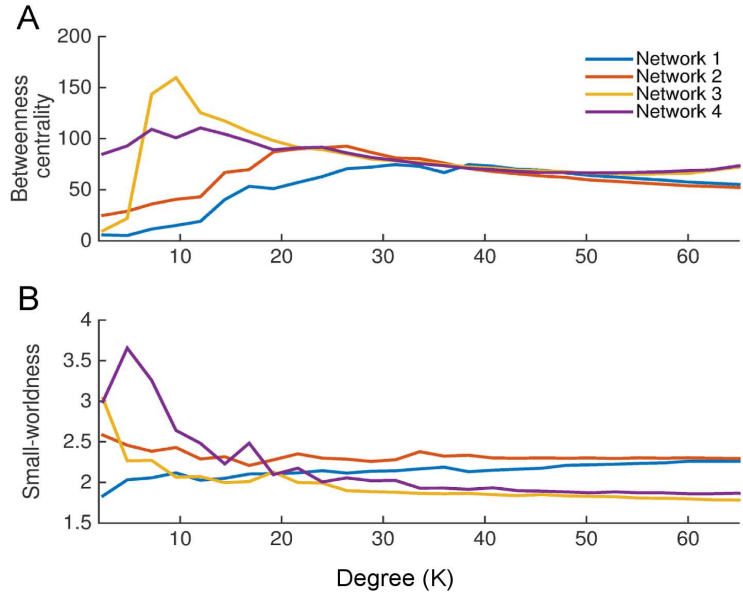}}
\caption{\textbf{Graph theoretical measures of the significant resting-state networks.} A) Node betweenness centrality measures (or hubs) as a function of the average node-degree measure $k$, B) small-worldness measure as a function of the average node-degree degree measure $k$.}
\label{fig5}
\end{figure}
\begin{figure}[htbp]
\centering{\includegraphics[height=9cm,width=10cm]{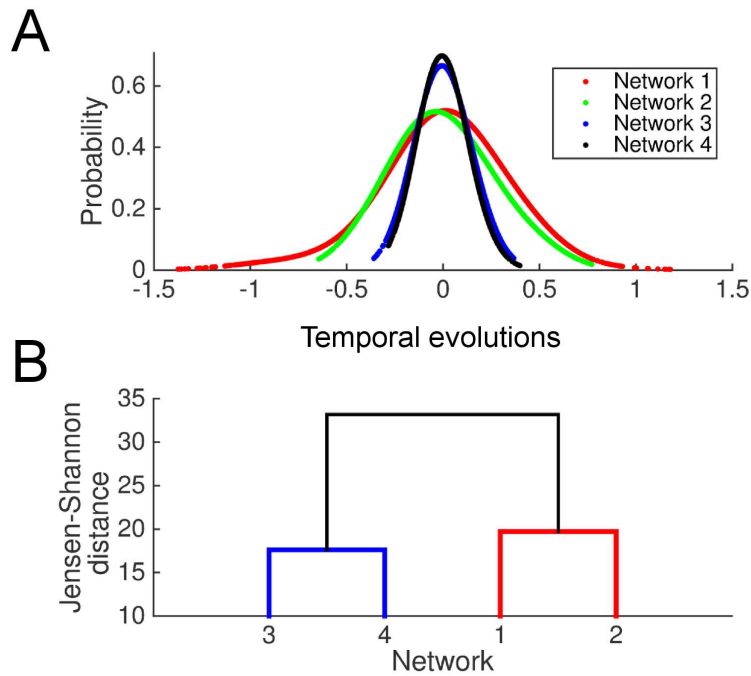}}
\caption{\textbf{Information theoretic properties of the temporal evolutions of significant resting-state networks.} A) Network clustering using Jensen-Shannon distance, B) Probability density functions of the temporal evolutions of the resting-state networks.}
\label{fig6}
\end{figure}
\subsection{Spectral-temporal contents of the resting-state networks}
\begin{figure}[htbp]
\centering{\includegraphics[height=15cm,width=14cm]{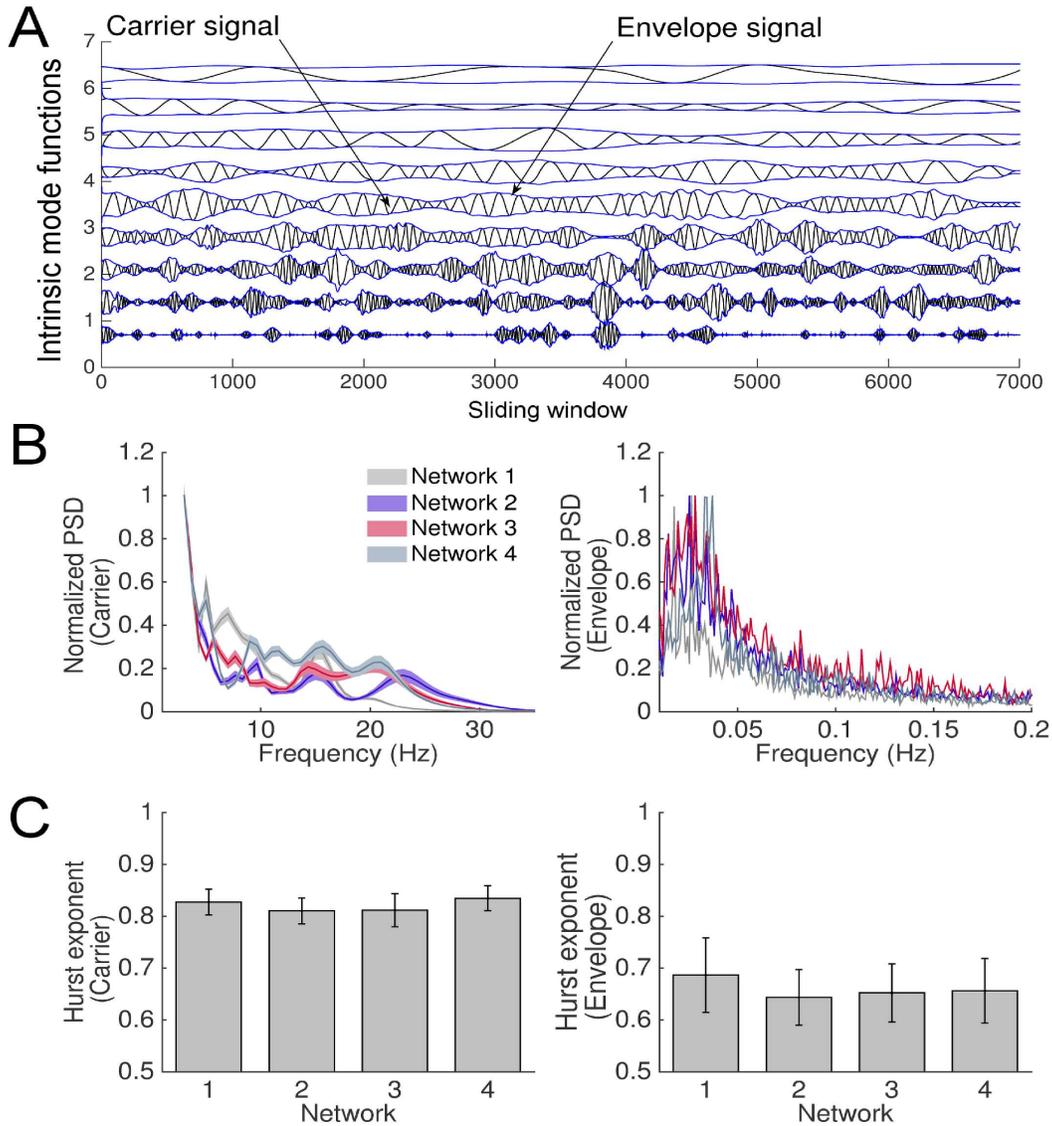}}
\caption{\textbf{Temporal-spectral contents of the resting-state networks}. A) The intrinsic mode functions derived from the temporal evolutions of network 1, B) log-power of the carrier signals derived from the first three IMFs of the 4 significant networks (left panel), log-power of the phase-coupling mode (i.e. envelope signals) derived from the first three IMFs of the 4 resting-state networks (right panel), C) the Hurst exponents of the carrier signals of the 4 significant networks (left panel), the Hurst exponents of the envelope signals of the 4 significant networks (left panel).}
\label{fig7}
\end{figure}

We also sought to characterise spectrum of the networks oscillations. To this end, a complex-valued Morlet-wavelet power spectra and the Hilbert amplitudes of the IMFs of each resting-state network were computed. The power spectra and the Hilbert amplitudes of the IMFs rendered distinct frequencies of the resting-state network oscillations $-$ the synchronous activity of large-scale neuronal populations (Figure \ref{fig7}). The IMFs captured distinct oscillatory patterns associated with each network (panel A). As expected, power spectra and the Hilbert amplitudes of the IMFs of each network revealed fast (rapid fluctuations underlying the envelope signal) and slow (envelope signal) neuronal activities associated with each network (panel B). Activity across all networks showed a broad spectral base indicating a broadband carrier frequency, although some networks have distinct spectral peaks superimposed on this base. Notably, all networks show a marked peak in the mu frequency (8$-$13 Hz) and beta frequency (18$-$25 Hz). In all networks, maximum power at the lower frequencies diminishes gradually towards higher frequencies, which is observed in both envelope and carrier signals/frequencies (panel B, left culumn). Notably, temporal evolutions of all networks exhibited highest power at frequencies well below 0.1 Hz (panel B, right column), the characteristic frequency of fMRI-derived resting-state networks, with an apparent 1/$f$-like distribution (i.e. power-law scaling) that is relatively invariant across the 4 significant networks. This effect indicates a strong interdependence of the amplitude and phase spectra such that the envelope of the networks' carrier frequencies are modulated across a broad spectrum of very slow timescales. Together, these spectra provide spectral-temporal fingerprints of the constraint dynamics of resting-state activity that is not well apparent in fMRI-derived data. As in previous study \citep{VanDeVille201018179}, we sought a more precise quantification of the apparent power-law scaling of the fast and slow network modulations by estimating the Hurst exponents. The average Hurst exponent of the IMFs derived from each network lie in the ranges of 0.8$-$0.9 for the carrier signals (panel C, left column), and 0.65$-$0.7 for the envelope signals (panel C, right column) respectively. Crucially, the confidence intervals of the Hurst exponent of all networks remain well above 0.5, which implies that the networks have persistent, long-range temporal dependencies, consistent with slow power-law decay. The Hurst exponents of the IMFs of the carrier signals (panel C, left column) were not significantly different (p=0.21) than the analogous Hurst exponents of the envelope frequencies (panel C, right column). 
\newpage
\section{Discussion}
In this study, we introduced a novel neurophysiology data-driven approach for  extracting spatial patterns of whole-brain FC with distinct topographies and spectral-temporal dynamics in resting brain. The main finding of this study was the existence of the patterns of rich-club organisation in resting-brain, by which the spectral and temporal variations of cortical activity are reshaped. Our technique applied SVD to large-scale brain FC of multiple subjects during resting-state. We found four significant large-scale connectivity patterns that encompassed regions from different functional clusters. The temporal evolutions (contributions) of the networks were obtained by taking the right-singular vectors from the SVD approach. By applying a kernel density approach and the EMD technique to the temporal evolutions of the networks, we found that the oscillatory activity of these networks were drawn from sub- and super- Gaussian distributions over multiple timescales, indicative of the existence of structured dynamics in coherent network variations. Specifically, the temporal contributions of these networks reflected slow modulation of fast oscillatory activity of cortical networks, which are consistent with other studies \citep{Hipp2012, Brookes201116783, Buzsaki20041926}.\\
\\
The spatial decomposition of fMRI data has become a widely employed means of identifying resting-state brain networks \citep{Beckmann20051001}. Yet the information in fMRI data is limited by the slow spatiotemporal hemodynamics \citep{Aquino201283}. Here we proposed a novel technique to reconstruct the low-dimensional subspace of cortical resting-state dynamics. Using source-reconstructed EEG data alone, we identified four significant resting-state networks that captured the dominant patterns of cortico-cortical interdependence across multiple subjects. These networks were constituted by a broadband carrier frequency $-$ revealed by the power spectra of the temporal evolutions associated with each network $-$ that is modulated by slow and persistent fluctuations revealed by the envelope spectra. The networks had distinctive spatial topographies, effectively spanning the `dynamic repertoire' \citep{Ghosh2008} of resting-state cortical activity. Whereas a variety of previous EEG and MEG studies have shown fast oscillatory activities that are correlated to slow time courses of networks identified by fMRI, the present technique captured low-dimensional resting-state dynamics from high-frequency neuronal synchronisation in electrophysiological data alone. The distinct patterns of brain-wide neuronal synchronisation and their spectral content that can be obtained with this method complements the building body of research on resting-state networks using BOLD fluctuations, band-pass filtered EEG and microstate analysis \citep{VanDeVille201018179, Hipp2012}. This approach provides a comprehensive picture of the dynamic patterns of neuronal synchronisation dynamics in resting-state cortical activity and supports the proposal that these networks arise through fast neuronal processes and not physiological noise or other structured artefacts in fMRI datasets, as may arise from non-gray matter, respiratory and cardiovascular signal fluctuation \citep{Birn20061536, Cole2010}.\\
\\
Our approach employs SVD to characterise the principled elements of dynamic FC and, consistent with previous studies, identifies prominent  changes in network topographies \citep{Hutchison2013360, Leonardi2013937, Allen2014663}. Our findings suggest that resting-state cortical networks are composed of slow modulations of fast coherent cortical activity. Because these modes arise in the absence of an explicit external task, they reflect the nature of organized cortical activity during spontaneous. They hence speak to the nature of candidate dynamical processes during spontaneous mental activity, including the role of nonlinear instabilities \citep{Honey2007fc}. Our findings, in particular the intrinsic mode functions of the temporal evolutions of the resting-state networks (Figure \ref{fig7}A), demonstrated how slow modulations (i.e. envelope signals) of fast cortical dynamics (i.e. carrier signals) can evolve dynamically at multiple timescales as revealed by the EMD approach. This effect is consistent with previous studies that found the activity of coherent cortical oscillations at multiple-timescales in resting-state \citep{Leonardi2013937, Mehrkanoon2014338}. Resting-state networks revealed transitions between different FC patterns: The time-course of FC obtained from the SVD revealed a 1/$f$-like distribution, indicating scale-free power fluctuations (Figure \ref{fig7}B). This signifies that the activity of resting-state cortical networks may be near a critical point, yielding long-range temporal correlations \citep{Linkenkaer-Hansen20011370}.\\
\\ 
We examined the resting-state cortical activity by investigating the intrinsic functional organisation of the resting-state networks using the graph metric, rich-club topology. Our approach captured a finite set of patterns related to resting-state network dynamics whose functional organisations were mainly located in the regions of the DMN, SN, and motor network. The DMN, which mainly consisted of the medial prefrontal cortex, and bilateral inferior parietal lobe, was first proposed by Raichle and colleagues based on the observation that these brain regions consumed higher energy during the resting-state compared with when performing tasks \citep{Raichle20071083}. The salience network, which consisted of the bilateral anterior insula and the ventrolateral prefrontal cortex (Figure \ref{fig2}), was first defined by Seeley and colleagues \citep{Seeley20072349}. The salience network is involved actively in bottom-up detection of salience events, and is hence central in switching between the brain's intrinsic large-scale functional organisation and networks in order to facilitate the access of resources from working memory \citep{Menon2010655}. The main role of the anterior insula was to activate the executive network and deactivate the DMN across occipital lobe, auditory cortex and in the resting-state as revealed by a measure of the effective connectivity, Granger causality analysis \citep{Sridharan200812569}.\\
\\
Interestingly, our findings showed that resting-state cortical networks were mainly associated with the activation and/or modulation of the specific hubs of the DMN, SN and motor network with distinct  complex topological features $-$ Rich-club organisations of the resting-state networks.  Surprisingly, the current study demonstrated that the dynamics of resting-state cortical activities were associated with changes in the fast-timescale spectral contents of the time-courses of the networks whilst they were modulated by slow-timescale. The four significant networks, each with distinct topographies and spectral-temporal contents, revealed a low-dimensional dynamical space by which resting-state cortical activities unfold. This implication is consistent with previous computational and empirical studies \citep{Aburn2012, Deco201143, Haimovici2013, VanDeVille201018179, Di201337}. Moreover, the Hurst exponents of the time-courses of the four resting-state networks were well above 0.5, which is also consistent with findings in previous study on resting EEG microstates \citep{VanDeVille201018179}. This indicates scale-free dynamics and complexity of resting-state cortical activity.\\
\\
By quantifying the complex topological features of the resting-state networks in  source-reconstructed EEG, we demonstrated that dynamic FC is mainly related to the activation of a finite set of brain networks such as DMN, SN, and motor network. Specifically, this arises the following questions, 1) Are patterns of network dynamics captured in the resting-state related to sequences of network states? 2) Is this a finite set of functional connectivity patterns supporting cognition, and is this repertoire continually revisited in the resting brain? \citep{Sporns2013162}. Further studies are required to address whether there is a global mechanism that governs a systematic interplay between the canonical brain networks (i.e. DMN, SN, and motor network) during different cognitive and motor tasks. This may reveal how the intrinsic large-scale brain dynamics is reshaped with respect to the connectivity configuration and a possible systematic interplay between different networks.\\ 
\newpage

\end{document}